\documentclass[pra,twocolumn,showpacs,amsmath,amssymb,superscriptaddress]{revtex4}

\usepackage{graphicx}
\usepackage{latexsym}
\usepackage{amsmath,amssymb}
\usepackage{verbatim,times,bbm}

\begin{document}

\newcommand{\beq}{\begin{equation}}
\newcommand{\eeq}{\end{equation}}
\newcommand{\barr}{\begin{eqnarray}}
\newcommand{\earr}{\end{eqnarray}}
\newcommand{\tr}{{\mathrm{tr}}}
\newtheorem{theorem}{Theorem}
\newtheorem{proposition}{Proposition}

\title{Gaussian maximally multipartite entangled states}
\author{Paolo Facchi}
\affiliation{Dipartimento di Matematica, Universit\`a di Bari,
        I-70125  Bari, Italy, EU}
\affiliation{INFN, Sezione di Bari, I-70126 Bari, Italy, EU}

\author{Giuseppe Florio}
\affiliation{Dipartimento di Fisica, Universit\`a di Bari,
        I-70126  Bari, Italy, EU}
\affiliation{INFN, Sezione di Bari, I-70126 Bari, Italy, EU}

\author{Cosmo Lupo}
\affiliation{Dipartimento di Fisica, Universit\`{a} di Camerino,
I-62032 Camerino, Italy, EU}

\author{Stefano Mancini}
\affiliation{Dipartimento di Fisica, Universit\`{a} di Camerino,
I-62032 Camerino, Italy, EU} \affiliation{INFN, Sezione di Perugia,
I-06123 Perugia, Italy, EU}

\author{Saverio Pascazio}
\affiliation{Dipartimento di Fisica, Universit\`a di Bari,
        I-70126  Bari, Italy, EU}
\affiliation{INFN, Sezione di Bari, I-70126 Bari, Italy, EU}

\begin{abstract}
We study maximally multipartite entangled states in the context of
Gaussian continuous variable quantum systems. By considering
multimode Gaussian states with constrained energy, we show that
perfect maximally multipartite entangled states, which exhibit the
maximum amount of bipartite entanglement for all bipartitions, only
exist for systems containing $n=2$ or 3 modes. We further
numerically investigate the structure of these states and their
frustration for $n\le 7$.
\end{abstract}


\maketitle

\section{Introduction}

Entanglement is nowadays recognized as a fundamental resource for
quantum information processing (see e.g.\ \cite{vlatko}). It
explicitly appeared long before the dawn of quantum information
science and without any reference to discrete variables (qubits)
\cite{Sch}. In fact, it first came to light in the context of
continuous variables \cite{EPR}. Thus, its characterization must
necessarily include the latter as well. Along this line, important
milestones have appeared in terms of continuous variables and more
specifically Gaussian states (see e.g.\ \cite{cvbooks} and reference
therein).

Although bipartite entanglement can be conveniently characterized
(e.g.\ in terms of purity or von Neumann entropy) \cite{entgen}, the
characterization of multipartite entanglement remains a challenging
problem, together with the definition of a class of quantum states
that exhibit high values of multipartite entanglement. Recently, the
notion of maximally multipartite entangled state (MMES) was
introduced in the qubit framework \cite{mmes}. These states have a
large (in fact, maximum) value of average bipartite entanglement
over all balanced bipartitions of a system of qubits
\cite{mmes,Scott}. They are solution of an optimization problem and
minimize  a suitably defined cost function, that can be viewed as a
potential of multipartite entanglement. A MMES is called ``perfect"
if its average entanglement saturates the maximum bipartite
entanglement for all bipartitions. Perfect MMESs exist for $n=2,3,5$
and 6 qubits, they do not exist for $n=4$, $n > 7$ \cite{lit,mmes},
while the case $n=7$ is still an open problem. In terms of potential
applications in quantum information science, MMESs are the ideal
resource for initializing a quantum internet \cite{Kimble} and could
be useful in several multiparty quantum information protocols (like
e.g.\ controlled teleportation \cite{Karl} or quantum secret sharing
\cite{Buz}).

The concept of MMES was extended to the framework of continuous
variable (and Gaussian) systems in \cite{adesso}. There a Gaussian
MMES is a state with a maximal rank for any bipartition of the $n$
party system in the limit of infinite squeezing \cite{adesso}.
Notice that such a state allows perfect quantum teleportation among
its $n$ parties. Here, with the aim of characterizing Gaussian
MMESs, we adopt a different viewpoint, by introducing a constraint
on the maximal mean energy allowed per user (which, eventually, will
be let to go to infinity). Hence, we look for states that present
the maximal amount of bipartite entanglement compatible with the
given constraint. We will show that, following this definition,
perfect MMESs exist in the continuous variable Gaussian setting only
for $n=2,3$. For $n \ge 4$ a simple argument shows that perfect
MMESs do not exist, hence manifesting the phenomenon of
\emph{entanglement frustration} \cite{frust} (see also
\cite{frust2}). Then, for $4 \le n \le 7$ we study the distribution
of entanglement among the bipartitions. Finally, we find examples of
MMESs and provide numerical evidence that bipartite entanglement can
be optimally distributed for $n=5,6$.

\section{Basic definitions} \label{basicdef}

A system composed of $n$ identical (but distinguishable) subsystems
is described by a Hilbert space $\mathcal{H}=\mathcal{H}_S$, with
$\mathcal{H}_S := \bigotimes_{i\in S} \mathfrak{h}_i$ and
$S=\{1,2,\dots,n\}$, which is the tensor product of the Hilbert
spaces of its elements $\mathfrak{h}_i\simeq \mathfrak{h}$. Examples
range from qubits, where $\mathfrak{h}=\mathbb{C}^2$, to continuous
variables systems, where $\mathfrak{h}=L^2(\mathbb{R})$. We will
denote a bipartition of system $S$ by the pair $(A,\bar{A})$, where
$A\subset S$, $\bar{A}= S \setminus A$ and $1\leq n_A \leq
n_{\bar{A}}$, with $n_A=|A|$, the cardinality of party $A$ (of
course, $n_A+n_{\bar{A}}=n$). At the level of Hilbert spaces we get
\begin{equation}
\mathcal{H}=\mathcal{H}_A \otimes \mathcal{H}_{\bar{A}}.
\end{equation}
A crucial question in quantum information  is about the amount of
entanglement between party $A$ and party $\bar{A}$. When the total
system is in a pure state $|\psi\rangle \in \mathcal{H}$, which is
the only case we will consider henceforth, the answer is simple, and
can be given, for example, in terms of the purity
\begin{equation}\label{eq:piAdef}
\pi_A = \tr (\rho_A^2)
\end{equation}
of the reduced density matrix of party $A$,
\begin{equation}
\rho_A := \tr_{\mathcal{H}_{\bar{A}}} (|\psi\rangle\langle\psi|) /
\|\psi\|^2.
\end{equation}
Indeed, this quantity can be taken as a measure of the entanglement
of the bipartition $(A,\bar{A})$. Its range is
\begin{equation}\label{eq:piArange}
\pi_{\mathrm{min}}^{n_A} \leq \pi_A \leq 1,
\end{equation}
where
\begin{equation}
\pi_{\mathrm{min}}^{n_A}=  (\dim \mathcal{H}_A)^{-1}=(\dim
\mathfrak{h})^{-n_A},
\end{equation}
with the stipulation that $1/\infty = 0$.

The upper bound $1$ is attained by unentangled, factorized states
$|\psi\rangle=|\phi\rangle_A\otimes |\chi\rangle_{\bar{A}}$.  On the
other hand, when $\dim \mathfrak{h}<\infty$, the lower bound, which
depends only on  the number of elements $n_A$ composing party $A$,
is attained by maximally bipartite entangled states, whose reduced
density matrix is a completely mixed state
\begin{equation}
\rho_A=\pi_{\mathrm{min}}^{n_A}  \openone_{\mathcal{H}_A}.
\end{equation}

Note, however, that for continuous variables, $\dim
\mathfrak{h}=\infty$, the lower bound $\pi_{\mathrm{min}}^{n_A}=0$
is not attained by any state. Therefore, strictly speaking, in this
situation  there do not exist maximally bipartite entangled states,
but only states that approximate them. This inconvenience can be
overcome by introducing physical constraints related to the limited
amount of resources that one has in real life. This reduces the set
of possible states and induces one to reformulate the question in
the form: what are the physical minimizers of (\ref{eq:piAdef}),
namely the states that minimize (\ref{eq:piAdef}) and belong to the
set $\mathcal{C}$ of physically constrained states? In sensible
situations, e.g.\ when one considers states with bounded energy and
bounded number of particles,  the purity lower bound
\begin{equation}
\pi_{\mathrm{min}}^{n_A, \mathcal{C}} = \inf\{\pi_A,
|\psi\rangle\in\mathcal{C} \}\geq \pi_{\mathrm{min}}^{n_A}
\end{equation}
is no longer zero and is attained by a class of minimizers, namely
the \emph{maximally bipartite entangled states}. If this is the
case, we can also consider multipartite entanglement and ask whether
there exist states in $\mathcal{C}$ that are maximally entangled for
every bipartition $(A,\bar{A})$, and therefore satisfy the extremal
property
\begin{equation}
\pi_A = \pi_{\mathrm{min}}^{n_A, \mathcal{C}}
\label{eq:perfectMMESdef}
\end{equation}
for every subsystem $A\subset S$ with $n_A=|A| \leq n/2$. In analogy
with the discrete variable situation, where $\dim
\mathfrak{h}<\infty$ and $\mathcal{C}=\mathcal{H}$, we will call a
state that satisfies (\ref{eq:perfectMMESdef}) a \emph{perfect MMES}
(subject to the constraint $\mathcal{C}$).

Since the requirement (\ref{eq:perfectMMESdef}) is very strong, the
answer to the quest can be negative for  $n>2$  (when $n=2$ it is
trivially satisfied) and the set of perfect MMES can be empty. We
remind that for a system of $n$ qubits, i.e.\ when $\dim
\mathfrak{h}=2$, perfect MMESs exist for $n=2,3,5,6$, do not exist
for $n=4$, $n > 7$ \cite{lit,mmes}, while the case $n=7$ is still
open. This is a symptom of frustration \cite{frust}. We emphasize
that this frustration is a consequence of the conflicting
requirements that entanglement be maximal for all possible
bipartitions of the system.

In the best of all possible worlds one can still seek for  the
(nonempty) class of states that better approximate perfect MMESs,
that is states with minimal average purity. We therefore consider
the \emph{potential of multipartite entanglement} \cite{mmes}
\begin{equation}
\label{eq:pimedef} \pi_{\mathrm{ME}} =  \mathbb{E}(\pi_A) :=
\left(\begin{array}{c}n
\\{[n/2]}\end{array}\!\!\right)^{-1}\sum_{|A|=[n/2]}\pi_A
\end{equation}
where the sum runs over all balanced bipartition, with $|A| = n_A =
[n/2]$, where $[\,\cdot\,]$ denotes the integer part. (It is
immediate to see that a necessary and sufficient condition for a
state to be a perfect MMES is to be maximally entangled with respect
to balanced bipartitions, i.e.\ those with $n_A=[n/2]$.) By
definition a MMES is a state that  belongs to $\mathcal{C}$ and
minimizes the potential of multipartite entanglement. Obviously,
when $\pi_{\mathrm{ME},\mathrm{min}}^\mathcal{C} :=
\min_{\mathcal{C}} \pi_{\mathrm{ME}}= \pi_{\mathrm{min}}^{[n/2],
\mathcal{C}}$ there is no frustration and the MMESs are perfect. In
order to quantify the amount of the frustration (for states
belonging to the set $\mathcal{C}$) we will take the quantity
\begin{eqnarray}\label{merit1}
\chi_{\mathrm{min}}^\mathcal{C} =
\pi_{\mathrm{ME},\mathrm{min}}^\mathcal{C}/
\pi_{\mathrm{min}}^{[n/2],\mathcal{C}}.
\end{eqnarray}
Eventually we will consider the limit
$\mathcal{C}\rightarrow\mathcal{H}$.

\section{Basic Tools for Gaussian states}

Let us consider a collection of $n$ identical bosonic oscillators
with (dimensionless) canonical variables $\{ q_k, p_k \}_{k=1,\dots
n}$. We assume that the oscillators have unit frequency and set
$\hbar=1$. A quantum state of the $n$ oscillators can be described
by a density operator $\rho_{(n)}$ on the $n$-mode Hilbert space, or
equivalently by the Wigner function on the $n$-mode phase space
\begin{equation}\label{Wigner}
W_{(n)}(\mathbf{q},\mathbf{p}) = \int d^n\mathbf{y} \langle
\mathbf{q} - \mathbf{y} | \rho_{(n)} | \mathbf{q} + \mathbf{y}
\rangle e^{2 i \pi \mathbf{y}\cdot\mathbf{p}},
\end{equation}
where $\mathbf{q} := (q_1, \dots q_n) $, $\mathbf{p} := (p_1, \dots
p_n) $, $\mathbf{y} := (y_1, \dots y_n) \in\mathbb{R}^n$, and we
have denoted by
\begin{equation}
| \mathbf{q} \pm \mathbf{y} \rangle := \otimes_{k=1}^n | q_k \pm y_k
\rangle
\end{equation}
the generalized eigenstates of the `position' operators $\hat q_k $.
By definition, Gaussian states are those described by a Gaussian
Wigner function. Introducing the phase-space coordinate vector
$\mathbf{X} = (X_1, \dots X_{2n}) := (q_1, p_1, \dots q_n, p_n)$, a
Gaussian state has a Wigner function of the following form:
\begin{eqnarray}
W_{(n)}(\mathbf{X}) &=& \frac{1}{(2\pi)^n\sqrt{\det(\mathbb{V})}}\nonumber\\
&\times&\exp{\left[-\frac{1}{2}(\mathbf{X}-\mathbf{X}_0)
\mathbb{V}^{-1}(\mathbf{X}-\mathbf{X}_0)^\mathsf{T}\right]},
\end{eqnarray}
where $\mathbf{X}_0 = \langle \mathbf{X} \rangle$, with $ \langle
f(\mathbf{X}) \rangle := \int f(\mathbf{X}) W_{(n)}(\mathbf{X})
d^{2n}\mathbf{X}$, is the vector of first moments, and $\mathbb{V}$
is the $2n\times 2n$ covariance matrix (CM), whose elements are
\begin{equation}
\mathbb{V}_{lm} = \langle ( X_l - \langle X_l \rangle )( X_m -
\langle X_m \rangle ) \rangle.
\end{equation}

We will also consider an equivalent representation defined by a
different ordering of the canonical variables $\mathbf{\tilde
X}=(q_1, q_2, \dots q_n, p_1, p_2 \dots p_n)$. In this
representation the CM is denoted $\mathbb{\tilde V}$ and has
elements
\begin{equation}\label{repr_2}
\mathbb{\tilde V}_{lm} = \langle ( \tilde X_l - \langle \tilde X_l
\rangle )( \tilde X_m - \langle \tilde X_m \rangle ) \rangle.
\end{equation}

In order to study the properties of entanglement for Gaussian states
we will consider the purity
\begin{equation}
\pi(\rho_{(n)}):=\tr(\rho_{(n)}^2).
\end{equation}
From Eq.~(\ref{Wigner}), it is straightforward to compute this
quantity in terms of the Wigner function:
\begin{equation}\label{eq:puritywigner}
\pi(\rho_{(n)})= (2\pi)^n\int \left[W_{(n)}(\mathbf{X})\right]^2
d^{2n}\mathbf{X}.
\end{equation}
In particular, the purity of Gaussian states is a function of the
determinant of the CM. From Eqs.~(\ref{eq:puritywigner}) it follows
that
\begin{equation}\label{purity}
\pi(\rho_{(n)}) = \frac{1}{2^n\sqrt{\det (\mathbb{V})}}
\end{equation}
with the bound $\pi(\rho_{(n)})  \le 1$. Notice that from
(\ref{purity}) a Gaussian state with positive CM is pure if and only
if
\begin{equation}\label{pur_cond}
\det \mathbb{V}  = \left(\frac{1}{2}\right)^{2n}.
\end{equation}

As anticipated in Sec. \ref{basicdef}, in order to obtain sensible
results, we will impose impose a suitable energy constraint. Here we
do not allow more than $N$ mean excitations for each bosonic mode,
i.e.\
\begin{equation}\label{energy}
\frac{\langle q_k^2 + p_k^2 \rangle}{2} \le N + \frac{1}{2}, \qquad
\mbox{for} \quad k=1, \dots n.
\end{equation}
This constraint introduces a cutoff in the Hilbert space of each
quantum oscillator.

A particular example of Gaussian state is the thermal state
$\rho_{(n)}^{\mathrm{th}}$ with $N$ thermal excitations per mode
described by a Gaussian Wigner function with vanishing first moments
and CM
\begin{equation}\label{V_thermal}
\mathbb{V}^{\mathrm{th}} = (N+1/2)\mathbb{I}_{2n}.
\end{equation}
Obviously, $\rho_{(n)}^{\mathrm{th}}$ satisfies the constraint
(\ref{energy}). We now show the following
\begin{proposition}\label{min_thermal}
Among all Gaussian states, the thermal state is the unique state
that minimizes purity under the constraint (\ref{energy}). The
corresponding minimal purity is
\begin{equation}\label{min_purity}
\pi_{\mathrm{min}}^{n,N}=\pi(\rho^{\mathrm{th}}_{(n)}) =
\frac{1}{2^n(N+1/2)^n}.
\end{equation}
\end{proposition}
\textbf{Proof:} We prove that the thermal state is the unique
minimizer of the purity among the Gaussian states satisfying the
inequality
\begin{equation}\label{energy_relax}
\frac{1}{n} \sum_{k=1}^n \frac{\langle q_k^2 + p_k^2 \rangle}{2} \le
N + \frac{1}{2},
\end{equation}
which constrains the average mean energy per mode. This is indeed
sufficient to prove the proposition since all the states satisfying
(\ref{energy}) also satisfy (\ref{energy_relax}).

The inequality (\ref{energy_relax}) can be written in terms of the
vector of first moments and the trace of the CM as follows
\begin{equation}\label{energy_relax_m}
\frac{\tr(\mathbb{V})+|\langle\mathbf{X}\rangle|^2}{2n} \le N +
\frac{1}{2}.
\end{equation}

Now we notice from Eq.\ (\ref{purity}) that the Gaussian state
minimizing the purity is the one whose CM has maximal determinant
under the constraint. The determinant and the trace of a CM are
functions of its eigenvalues $\{ v_j \}_{j=1,\dots 2n}$. We hence
consider the problem of finding $\mathbb{V}$ such that
$\det{(\mathbb{V})}=\prod_{j} v_j$ is maximal under the constraint
$\tr{(\mathbb{V})}=\sum_j v_j \le 2n(N+1/2)-|\langle \mathbf{X}
\rangle|^2$. The unique solution is the scalar matrix $[N+1/2 -
|\langle \mathbf{X} \rangle|^2/(2n)]\mathbb{I}_{2n}$. For a given
value of $\langle \mathbf{X} \rangle$ the maximal value of the
determinant is hence $[N+1/2 - |\langle \mathbf{X}
\rangle|^2/(2n)]^{2n}$. It follows that the unique Gaussian state
minimizing the purity under the energy constraint
(\ref{energy_relax}) --- and hence the constraint (\ref{energy})
--- is the one with $\langle \mathbf{X} \rangle = 0$ and CM as in
Eq.\ (\ref{V_thermal}), i.e.\ the thermal state. $\mathbf{QED}$

\section{Gaussian MMES}

In the following, we will focus our attention on the case of {\it
pure} Gaussian states characterized by Eq.\ (\ref{pur_cond}) and
subjected to the energy constraint (\ref{energy}). As in
Sec.~\ref{basicdef}, we consider a collection of $n$-modes and a
bipartition into two disjoint subsets $A$ and $\bar{A}$, containing
$n_A$ and $n_{\bar{A}}=n-n_A$ modes, respectively, with $n_A \le
n_{\bar{A}}$. In order to quantify the bipartite entanglement
between the two subsets of oscillators, we compute the purity
$\pi_A$ of the reduced state of subsystem $A$. The modes of subset
$A$ and $\bar{A}$ have phase-space coordinates $\mathbf{X}_A$ and
$\mathbf{X}_{\bar{A}}$, respectively. The Wigner function describing
the reduced state of party $A$ is obtained by integrating the Wigner
function of the whole system over the variables belonging to
$\bar{A}$, i.e.\
\begin{equation}
W_{(n_A)}(\mathbf{X}_A) = \int W_{(n)}(\mathbf{X})\; d^{2
n_{\bar{A}}} \mathbf{X}_{\bar{A}}.
\end{equation}
It follows that the reduced state of a Gaussian state with first
moment $\langle \mathbf{X} \rangle$ and CM $\mathbb{V}$ is Gaussian
{with first moment $\langle \mathbf{X}_A \rangle$ and} CM
$\mathbb{V}_{A}$. The CM of the reduced state is the square
sub-matrix of $\mathbb{V}$ identified by the indices belonging to
subsystem $A$.

States that are maximally entangled with respect to the given
bipartition are those admitting a reduced state for subsystem $A$
with minimum value for the purity. From Proposition
\ref{min_thermal} the reduced system has to be in a thermal state of
$n_A$ oscillators. Taking into account the energy constraint
(\ref{energy}) we get
\begin{equation}
\mathcal{C} = \left\{|\psi\rangle \in \mathcal{H},\; |\psi\rangle
\text{ Gaussian},\; \frac{\langle q_k^2 + p_k^2 \rangle}{2} \le N +
\frac{1}{2} \right\} , \label{eq:physconstr}
\end{equation}
and
\begin{equation}
\label{eq:purityconstr} \pi_{\mathrm{min}}^{n_A,\mathcal{C}}\leq
\pi_A \leq 1
\end{equation}
with
$\pi_{\mathrm{min}}^{n_A,\mathcal{C}}=\pi_{\mathrm{min}}^{n_A,N}$
given by (\ref{min_purity}).

We will generalize this property to multipartite entanglement by
requiring minimal possible purity for \emph{each} subsystem $A$ of
the modes --- assuming that the state of the total system is pure
--- thus defining a \emph{Gaussian maximally multipartite entangled
state}. In particular, we will define a {\it perfect} Gaussian MMES
as a pure Gaussian state of $n$ oscillators that is maximally
entangled with respect to all balanced bipartitions and satisfies
the energy constraint (\ref {eq:physconstr}). It follows from this
definition that the reduced state is thermal for all possible
bipartitions $(A,\bar{A})$.

In order to formalize the above definition of MMES for Gaussian
states with constrained energy we start from (\ref{eq:pimedef}) and define a
(normalized) potential of multipartite entanglement:
\begin{eqnarray}\label{merit}
\chi := \pi_{\mathrm{ME}}/ \pi_{\mathrm{min}}^{n_A,N} =
(N+1/2)^{n_A} \mathbb{E}\left[\det(\mathbb{V}_A)^{-1/2}\right] ,
\end{eqnarray}
where $n_A = [n/2]$, $\mathbb{V}_A$ is the square sub-matrix defined
by the corresponding indexes. The minimum of this quantity
$\chi_{\mathrm{min}}^\mathcal{C}$ (where $\mathcal{C} \simeq N$ is
the constraint) is a measure of frustration, according to Eq.\
(\ref{merit1}). A Gaussian MMES will be a minimizer of the potential
(\ref{merit}). The potential in (\ref{merit}) is the normalized
purity of the reduced state, averaged over all balanced
bipartitions. Notice that $\chi \ge 1$, and perfect Gaussian MMESs
satisfy $\chi = 1$.

We recall from the above discussion that the requirement of the
minimization of purity for a given bipartition could be in contrast
with that for another bipartition. Therefore perfect MMESs do not
necessarily exist. Actually, we get the following
\begin{theorem}
Perfect Gaussian MMESs only exist for $n<4$.
\end{theorem}
\textbf{Proof.} First we present examples of Gaussian MMESs for
$n=2,3$, then we show that Gaussian MMESs do not exist for $n \ge
4$.

A two-mode Gaussian state is described by the vector of first
moments $\langle\mathbf{X}\rangle=\langle (q_1, p_1, q_2, p_2)
\rangle$ and by the $4 \times 4$ CM
\begin{eqnarray}
\mathbb{V} = \left(\begin{array}{cc} \mathbb{V}_{1,1} & \mathbb{V}_{1,2} \\
\mathbb{V}_{1,2}^\mathsf{T} & \mathbb{V}_{2,2}
\end{array}\right).
\end{eqnarray}
Imposing that the one-mode reduced states are thermal implies
$\langle\mathbf{X}\rangle=(0, 0, 0, 0)$ and
$\mathbb{V}_{1,1}=\mathbb{V}_{2,2}=(N+1/2)\mathbb{I}_2$. It remains
to specify the submatrix $\mathbb{V}_{1,2}$ in order to obtain a
well defined CM $\mathbb{V}$ satisfying the purity condition
(\ref{pur_cond}). A solution is given by the CM
\begin{eqnarray}
\mathbb{V} = \frac{1}{2}\left(\begin{array}{cccc}
\cosh{r} & 0         & \sinh{r} & 0 \\
0        & \cosh{r}  & 0        & -\sinh{r} \\
\sinh{r} & 0         & \cosh{r} & 0 \\
0        & -\sinh{r} & 0        & \cosh{r}
\end{array}\right),
\end{eqnarray}
describing a two-mode squeezed state, the so-called twin-beam state
\cite{cvbooks}, for
\begin{equation}
\cosh{r} = 2N+1
\end{equation}
(squeezing parameter $r/2$).

Let us now consider the case $n=3$. A three-mode Gaussian MMES has
all the three single-mode reduced systems in a thermal state. Hence
the vector of first moments vanishes, and the CM has the form
\begin{eqnarray}
\mathbb{V} = \left(\begin{array}{ccc}
(N+1/2)\mathbb{I}_2 & \mathbb{V}_{1,2} & \mathbb{V}_{1,3} \\
\mathbb{V}_{1,2}^\mathsf{T} & (N+1/2)\mathbb{I}_2 & \mathbb{V}_{2,3} \\
\mathbb{V}_{1,3}^\mathsf{T} & \mathbb{V}_{2,3}^\mathsf{T} &
(N+1/2)\mathbb{I}_2 \end{array}\right).
\end{eqnarray}
It remains to determine the sub-matrices $\mathbb{V}_{1,2}$,
$\mathbb{V}_{1,3}$, $\mathbb{V}_{2,3}$ in order to obtain a well
defined CM obeying the purity constrain (\ref{pur_cond}). A solution
is given by an instance of the tripartite Gaussian GHZ states
\cite{GHZ} characterized by the condition
$\mathbb{V}_{1,2}=\mathbb{V}_{1,3}=\mathbb{V}_{2,3}=\mathrm{diag}(v_+,v_-)$
with
\begin{equation}
v_\pm = \frac{N(N+1)}{4N+2}\left[ 1 \pm \sqrt{ 1 +
\frac{(4N+2)^2}{2N(N+1)} } \right].
\end{equation}

The cases $n=2,3$ are the only ones in which perfect Gaussian MMESs
exist. The non existence of perfect Gaussian MMESs for $n \ge 4$ is
easily seen by inspecting the $n$-mode CM. Indeed, the generic
submatrix of $n_A \le n/2$ modes is of the form
\begin{eqnarray}
\mathbb{V}_A = \left(\begin{array}{lll}
\mathbb{V}_{i_1,i_1}& \ldots & \mathbb{V}_{i_1,i_{n_A}} \\
\vdots & \ddots & \vdots   \\
\mathbb{V}_{i_1,i_{n_A}}^\mathsf{T}  & \ldots & \mathbb{V}_{i_{n_A},i_{n_A}}
 \end{array}\right)
\end{eqnarray}
where $\mathbb{V}_A$ is a $2n_A\times 2n_A$ matrix and
$A=\{i_1,\dots,i_{n_A}\}$. For $n \ge 4$, the definition of perfect
Gaussian MMES implies that
\begin{eqnarray}
\mathbb{V}_{i,j}&=&0 \quad \mbox{for}\,\, i,j\in A\,\,\mbox{and}\,\,  i\ne j,\\
 \mathbb{V}_{i,i}&=&(N+1/2)\mathbb{I}_{2} \quad \mbox{for}\,\, i\in A.
\end{eqnarray}
This condition must hold for all bipartitions $(A,\bar{A})$ and,
therefore, all off-diagonal sub-matrices are zero. As a consequence,
the CM of the Gaussian state is diagonal of the form
$(N+1/2)\mathbb{I}_{2n}$. Such a CM describes a thermal state with
$N$ thermal excitation per mode, in contradiction with the
requirement that the global state of the $n$ oscillators is pure.
$\mathbf{QED}$

\section{Numerical search of MMES}

We have seen that perfect MMESs only exist for $n < 4$. For $n \ge
4$ we now numerically search for $n$-mode pure states minimizing the
cost function (\ref{merit}), under the energy constraint
(\ref{energy}). Minimizing the cost function (\ref{merit})
corresponds to minimizing the average purity of the reduced states.
The value $\chi=1$ corresponds to a perfect MMES. For $n\ge 4$ this
is possible only for $N=0$, where the problem becomes trivial since
the only state compatible with the energy constraint is the vacuum,
which is a separable state.

For numerical investigations we use a convenient parametrization of
$n$-mode pure Gaussian states. First of all in the following we will
assume, without loss of generality, $\langle \mathbf{X} \rangle =
\langle \mathbf{\tilde X} \rangle = 0$. It remains to parameterize
the set of $n$-mode covariance matrices. Working in the
representation (\ref{repr_2}), it is possible to show that the CM of
a $n$-mode pure Gaussian state can be written as \cite{deGo}
\begin{equation}
\mathbb{\tilde V} = \frac{1}{2} \mathbb{R} \mathbb{T}^2
\mathbb{R}^\mathsf{T},
\end{equation}
where $\mathbb{T}$ is a diagonal matrix of the form
\begin{eqnarray}
\mathbb{T} = \left(\begin{array}{cc} \mathbb{K} & 0 \\
0 & \mathbb{K}^{-1} \end{array}\right),
\end{eqnarray}
with $\mathbb{K}$ a non singular diagonal matrix, while $\mathbb{R}$
is both symplectic and orthogonal. Therefore it has the form
\begin{eqnarray}
\mathbb{R} = \left(\begin{array}{cc} \mathbb{X} & \mathbb{Y} \\
-\mathbb{Y} & \mathbb{X} \end{array}\right),
\end{eqnarray}
where the matrix $\mathbb{U}=\mathbb{X}+i\mathbb{Y}$ is unitary.

\begin{figure}
\centering
\includegraphics[width=0.5\textwidth]{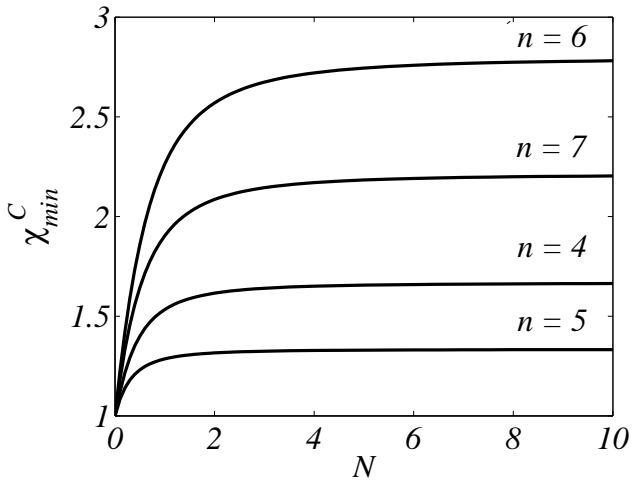}
\caption{Minimal value of the (dimensionless) cost function $\chi$
versus the (dimensionless) energy constraint parameter $N$, for
several values of $n$.} \label{chi_N}
\end{figure}

Figure \ref{chi_N} shows minimal value
$\chi_{\mathrm{min}}^\mathcal{C}$ of the potential (\ref{merit})
under the constraint (\ref{energy}), for $4\le n\le 7$, as a
function of the mean number of excitations per mode $N$. This
minimal value yields a measure of the {\it frustration} present in
the system, which does not allow the existence of perfect MMESs. The
larger the minimal value of $\chi$, the larger the frustration. The
numerical analysis indicates that the minimum of the potential of
multipartite entanglement is a nondecreasing concave function of
$N$; moreover, a plateau is reached for sufficiently high values of
$N$. This saturation value increases with $n$, but oscillates
between even and odd $n$.

Since it is not possible to find perfect MMESs, it is important to
quantify the distribution of entanglement. A good distribution of
entanglement should be rather insensitive to a change of
bipartition. A fairly distributed multipartite entanglement should
therefore be characterized by a distribution (over balanced
bipartition) with a small standard deviation \cite{FFP}. We
therefore consider the standard deviation of the purity of the
reduced states over balanced bipartitions
\begin{equation}\label{variance}
\Delta\chi := \sqrt{\left(N+1/2\right)^{2n_A}
\mathbb{E}\left[\det(\mathbb{V}_A)^{-1}\right] - \chi^2},
\end{equation}
with $n_A=[n/2]$. We will call a MMES with $\Delta\chi=0$ a
\emph{uniformly optimal MMES}, because it has an optimal
distribution of entanglement: entanglement (and frustration) is
fairly distributed over all bipartitions that attain the minimal
value of purity allowed by frustration. Of course, a perfect MMES is
uniformly optimal.

\begin{figure}
\centering
\includegraphics[width=0.5\textwidth]{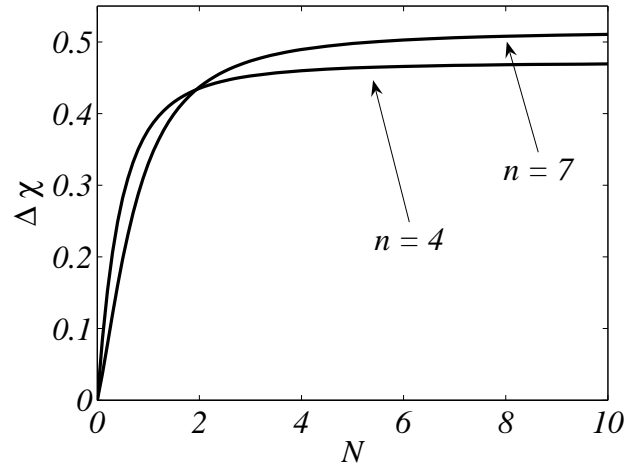}
\caption{Standard deviation $\Delta\chi$ (dimensionless) of the
normalized purity versus the (dimensionless) energy constraint
parameter $N$, for $n=4$ and $7$. For $n=5$ and $6$ the MMESs have
optimally distributed entanglement, hence vanishing standard
deviation.} \label{Dchi_N}
\end{figure}

Figure \ref{Dchi_N} displays the peculiar behavior of the standard
deviation of the purity: such standard deviation has a different
behavior as a function of $N$ for different values of $n$. For $n=5$
and $6$, MMESs are not perfect. Nonetheless, interestingly enough,
they have an optimal distribution of entanglement: $\Delta\chi=0$
(for $n=5,6$). By contrast, a non optimal entanglement distribution
has been found for MMES with $n=4$ and $7$. This reminds, {\it
mutatis mutandis}, of what happens in qubit systems, where
frustration appears for $n=4$ and $n \ge 7$ \cite{mmes,Scott,lit}.
The numerical analysis shows a $\Delta\chi$ of the non uniformly
optimal states which is a concave nondecreasing function of the
energy parameter $N$. We notice also in this case the presence of a
saturation effect for large values of $N$. These findings are
summarized in Table \ref{tab_qGmmes}.

\begin{table}[t]
\caption{Comparison between qubit and Gaussian maximally
multipartite entangled states.} \label{tab_qGmmes}
\begin{tabular}{|c|c|c|}
\hline
$n$ & qubit perfect MMES & Gaussian perfect MMES \\
\hline
    2,3 & yes  & yes \\
    4 & no & no  \\
     5,6 & yes & no, but uniformly optimal$^*$\\
     7 & $\;\,$no$^*$ & no \\
    $\geq 8$ & no & no  \\
    \hline
\end{tabular}
\newline
$^*$numerical evidence
\end{table}

\section{Conclusions}

In conclusion, we have characterized Gaussian states that display a
maximal amount of multipartite entanglement compatible with a given
constraint on the mean energy. We have shown that perfect Gaussian
MMESs (that saturate the maximum mean energy) only exist for $n=2$
and $3$, while the phenomenon of entanglement frustration appears
already for $n \ge 4$. Curiously, we found clear numerical evidence
that although perfect Gaussian MMESs do not exist for $n=5$ and $6$,
for these particular values of $n$, bipartite entanglement can be
optimally distributed, in the sense that the standard deviation of
purity over balanced bipartitions (\ref{variance}) can be made to
vanish. We numerically found that, by contrast, such standard
deviation \emph{cannot} be made to vanish (and bipartite
entanglement is therefore not optimally distributed) for $n=4$ and
$7$. This peculiar situation is reminiscent of that encountered with
qubit MMES, where perfect MMESs exist for $n=5$ and $6$ (qubits),
but do not exist for $n=4$ and (probably \cite{mmes}) $n=7$. This
suggests once more that $n=2,3,5$ and $6$ are ``special" integers.
We endeavored to summarize these conclusions in Table
\ref{tab_qGmmes}. Experience with integers does not induce us to
expect that these amusing peculiarities only occur for $n \leq 6$
(for instance, we numerically found that uniformly optimal MMESs ---
with vanishing $\Delta\chi$ --- exist for $n=9$). Additional
research is needed in order to investigate the large $n$ behavior
and clarify the underlying structure of entanglement frustration. We
emphasize again that this frustration is a consequence of the
conflicting requirements that entanglement be maximal for all
possible bipartitions of the system. The same phenomenon is also
worth studying under different constraints, like, for instance, the
(weaker) energy constraint (\ref{energy_relax}).

From a more applicative perspective, we emphasize that due to recent
progress in the optical generation of Gaussian entangled states (up
to $9$ modes \cite{Su}) the above features are also liable to
experimental check. These results, combined with those obtained in
\cite{adesso}, and the ensuing proposed characterization of
entanglement would help in optimizing multiparty quantum information
protocols with continuous variables.

\subsection*{Acknowledgments}

The work of PF, GF and SP is partly supported by the EU through the
Integrated Project EuroSQIP. The work of CL and SM is supported by
EU through the FET-Open Project HIP (FP7-ICT-221899). The authors
thank an anonymous referee for an important remark.

\end{document}